\documentclass[aps,preprintnumbers,nofootinbib,prd,superscriptaddress,tightenlines,notitlepage,twocolumn,showpacs,floatfix]{revtex4-1}
\usepackage[colorlinks]{hyperref}
\usepackage{tabularx}
\usepackage{graphicx}
\usepackage{amssymb,bm,tensor}
\usepackage{textcomp}
\usepackage{xcolor}
\usepackage{amsmath}
\usepackage[varg]{txfonts}
\usepackage{enumerate}
\usepackage{mathtools}
\usepackage[normalem]{ulem}
\usepackage{bm} 
\usepackage[OT1]{fontenc}
\usepackage{aas_macros}

\newcommand{\hata}{\hat{a}}
\newcommand{\hatb}{\hat{b}}
\newcommand{\hatc}{\hat{c}}
\newcommand{\hatn}{\hat{n}}
\newcommand{\JRA}[4]{$#1^{\rm h}#2^{\rm m}#3.\!^{\rm s}#4$}
\newcommand{\Jdec}[4]{$#1^\circ#2'#3.\!''#4$}

\newcommand{\KIAA}{\affiliation{Kavli Institute for Astronomy and Astrophysics,
Peking University, Beijing 100871, China}}
\newcommand{\MPIfR}{\affiliation{Max-Planck-Institut f\"ur Radioastronomie, Auf
dem H\"ugel 69, D-53121 Bonn, Germany}}
\newcommand{\ERAU}{\affiliation{Department of Physics and Astronomy, Embry-Riddle Aeronautical
University, Prescott, Arizona 86301, USA}}

\begin{document}

\title{Testing velocity-dependent ${\cal CPT}$-violating gravitational forces
with radio pulsars}
\date{\today}
\author{Lijing Shao}\email{lshao@pku.edu.cn}\KIAA\MPIfR
\author{Quentin G. Bailey}\email{baileyq@erau.edu}\ERAU
\preprint{MITP/18-086}

\begin{abstract}
    In the spirit of effective field theory, the Standard-Model Extension (SME)
    provides a comprehensive framework to systematically probe the possibility
    of Lorentz/CPT violation.  In the pure gravity sector, operators with mass
    dimension larger than 4, while in general being advantageous to short-range
    experiments, are hard to investigate with systems of astronomical size.
    However, there is exception if the leading-order effects are CPT-violating
    and velocity-dependent. Here we study the lowest-order operators in the
    pure gravity sector that violate the CPT symmetry with carefully chosen
    relativistic binary pulsar systems.  Applying the existing analytical
    results to the dynamics of a binary orbit, we put constraints on various
    coefficients for Lorentz/CPT violation with mass dimension 5.  These
    constraints, being derived from the post-Newtonian dynamics for the first
    time, are complementary to those obtained from the kinematics in the
    propagation of gravitational waves.
\end{abstract}

\maketitle


\section{Introduction}
\label{sec:intro}

There is a great deal of theoretical interest to probe new physics beyond the
Standard Model of particle physics, and the General Relativity (GR) theory of
gravitation~\cite{Goenner:2004se, Liberati:2013xla, Tasson:2014dfa,
Will:2014bqa}.  Most of them stem from the need for a theory of quantum
gravity, namely, to unify quantum field theories and GR, or in other words, to
describe the four fundamental forces within a single mathematical
setting~\cite{Kostelecky:1988zi, Burgess:2003jk, Hossenfelder:2012jw}.  Up to
now, although there are achievements at different levels, not one proposal has been
singled out as the widely accepted final theory for quantum gravity. On the
other hand, observational evidence that was accumulated during the past decades
--- with intriguing puzzles from dark matter, dark energy, and inflationary
cosmology, just to name a few --- points to the need going beyond the current
paradigm of modern theoretical physics~\cite{Clifton:2011jh, Will:2014kxa,
Tasson:2016xib}.

Broadly speaking, there are two ways to investigate new physics beyond our
current understanding: theory specific and theory agnostic. Effective field
theory (EFT) is a natural candidate framework for the
latter~\cite{Weinberg:2009bg, Burgess:2003jk}. In the spirit of EFT,
Kosteleck\'y and collaborators have developed a comprehensive framework, dubbed
the Standard-Model Extension (SME), to catalogue all possible operators that
are gauge invariant, Lorentz covariant, and energy-momentum
conserving~\cite{Kostelecky:1988zi, Colladay:1996iz, Colladay:1998fq,
Kostelecky:2003fs, Bailey:2006fd, Kostelecky:2010ze, Kostelecky:2017zob}. In
general, a violation in CPT implies a violation in the Lorentz
symmetry~\cite{Greenberg:2002uu}. In a practical way, we will collectively call
the coefficients of new operators beyond the Standard Model and GR {\it
coefficients for Lorentz/CPT violation}~\cite{Kostelecky:2008ts}.  During the
past decades, the SME has been successfully applied in various experiments, and
many constraints were set on the coefficients for Lorentz/CPT
violation~\cite{Kostelecky:2008ts, Hees:2016lyw, Shao:2016ezh}. No statistically convincing violation
has been found yet \cite{Kostelecky:2008ts}.

We here focus on the pure gravity sector of SME~\cite{Kostelecky:2003fs,
Bailey:2006fd, Kostelecky:2016uex, Shao:2016cjk, Bailey:2016ezm,
Bailey:2017lbo, Kostelecky:2017zob}. The general framework for Riemann-Cartan spacetime 
was described in Ref.\ \cite{Kostelecky:2003fs}.  To be mathematically compatible with the
Riemann-Cartan geometry, Lorentz/CPT breaking can be considered to be {\it spontaneous}, 
instead of {\it explicit} \cite{Bluhm:2015}. 
Extra dynamical fields in the framework
obtain their vacuum expectation values through symmetry breaking
cosmologically, in analog with the Higgs mechanism in the Standard Model.
However in SME these fields are not necessarily to be scalar fields, but can
take on nontrivial spacetime indices and therefore have tensorial nature.
Therefore, after symmetry breaking, the effective Lagrangian is {\it observer
Lorentz invariant}, but {\it particle Lorentz
violating}~\cite{Kostelecky:2003fs, Bailey:2006fd, Tasson:2016xib}.  To be
fully compatible with geometrical requirements at desired orders, the
underlying fluctuating Nambu-Goldstone modes that arise from the symmetry
breaking need to be propoerly accounted for~\cite{Kostelecky:2003fs,
Bailey:2006fd}.  In Ref.\ \cite{Bailey:2006fd} the post-Newtonian
behaviours from the pure-gravity sector of SME for operators with mass
dimension up to 4 were studied.  The leading-order post-Newtonian effects are described by a
tensor field, $\bar{s}^{\mu\nu}$, where the ``bar'' indicates that it is the
vacuum expectation value of the underlying dynamical field $s^{\mu\nu}$.
Different experiments, including lunar laser ranging~\cite{Battat:2007uh,
Bourgoin:2017fpo}, atom interferometers~\cite{Muller:2007es, Chung:2009rm,
Flowers:2016ctv}, cosmic rays~\cite{Kostelecky:2015dpa}, pulsar
timing~\cite{Shao:2012eg, Shao:2013wga, Shao:2014oha, Shao:2014bfa,
Xie:2013iia, Shao:2016dtg}, planetary orbital dynamics~\cite{Hees:2015mga}, and
gravitational waves~\cite{TheLIGOScientific:2017qsa, Monitor:2017mdv} were used
to constrain $\bar{s}^{\mu\nu}$ (see \citet{Hees:2016lyw} for a review).

Recently, higher-dimensional operators with mass dimension larger than 4
in the gravity sector of SME were investigated, and short-range gravity
experiments in laboratory were identified to be the best to constrain these
terms due to the extra powers in $1/r$ for the gravitational forces derived from
these operators~\cite{Bailey:2014bta, Shao:2016cjk, Shao:2016jzh,
Kostelecky:2016uex}. However, there is an exception. 
\citet{Bailey:2017lbo} found that the leading-order CPT-violating operators
with mass dimension 5 produce a gravitational force, between two objects $a$
and $b$, proportional to $\left( \bm{v}_a - \bm{v}_b \right)/r^3$.  For
short-range gravity experiments, $\left( \bm{v}_a - \bm{v}_b \right)/c$ is very
close to zero, thus these experiments are very hard, if ever possible, to probe
these terms.  Estimated sensitivities of different
experiments to these new operators were tabulated (see Table III in
Ref.~\cite{Bailey:2017lbo}), where binary pulsars turn out to be among the most
sensitive probes.  This motivates us to take a closer look at these new
operators, and to collect the best binary pulsars in order to derive
constraints on the coefficients for Lorentz/CPT violation.

The paper is organized as follows. In the next section, we review the structure
of the gravity sector of SME at leading orders, and give the expressions for
secular changes for elements of a binary orbit~\cite{Bailey:2006fd,
Bailey:2017lbo}. Then in section~\ref{sec:psr} we carefully choose the binary
pulsars that are suitable for the test, and discuss our approach to evade
difficulties related to observationally unknown angles and the consistency in
using timing parameters with {\it a priori} unknown component masses.  Our
direct constraints are summarised in Table~\ref{tab:Ki}, and they are properly
converted to constraints on the coefficients in the Lagrangian in
Tables~\ref{tab:single} and \ref{tab:global}.  In the last section we point out
the perturbative nature of SME and the post-Newtonian approach, thus we should
keep caveats in mind when dealing with strongly self-gravitating bodies like
neutron stars (NSs)~\cite{Damour:1993hw, Shao:2017gwu}.  Throughout the paper, unless
explicitly stated, we use units where $\hbar=c=1$.

\section{Theory} \label{sec:theory}

At present there are two approaches to the gravity sector of the SME. The first
is a general coordinate invariant version~\cite{Kostelecky:2003fs}, while the
second focuses on a spacetime that can be expanded around a Minkowski
metric~\cite{Kostelecky:2017zob}.  These two approaches have distinct
underlying methodology, but are interrelated.  We use the latter in this work.
We restrict ourselves to the discussion of the part of spacetime where, after
fixing the gauge (say, the harmonic gauge), linearized gravity is a good
approximation.  The metric is decomposed into a flat-spacetime metric,
$\eta_{\mu\nu} \equiv {\rm diag} \left\{ -,+,+,+ \right\}$, and a perturbation,
$h_{\mu\nu}$,
\begin{align}
    g_{\mu\nu} = \eta_{\mu\nu} + h_{\mu\nu} \,,
\end{align}
where $\left| h_{\mu\nu} \right| \ll 1$. With this assumption, it is possible
to write down the generic Lagrangian density for a spin-2 massless particle,
organized by the order of the mass dimension of the coupling coefficients
~\cite{Kostelecky:2003fs, Bailey:2006fd, Bailey:2014bta, Kostelecky:2016kfm,
Bailey:2017lbo},
\begin{align}\label{eq:lagrangian}
    {\cal L} &= {\cal L}_{\rm GR} + {\cal L}_{\rm SME}^{(4)} + {\cal L}_{\rm
    SME}^{(5)} + \cdots \,,
\end{align}
where the GR terms are,
\begin{align}
    {\cal L}_{\rm GR} &= -\frac{1}{32\pi G} h^{\mu\nu} G_{\mu\nu} + \frac{1}{2}
    h_{\mu\nu} T^{\mu\nu}_{\rm matter} \,,
\end{align}
with $G_{\mu\nu}$ the linearized Einstein tensor, and $T^{\mu\nu}_{\rm matter}$
the matters' energy-momentum tensor.  

The leading-order corrections in Eq.~(\ref{eq:lagrangian})
are~\cite{Bailey:2017lbo},
\begin{align}
    {\cal L}_{\rm SME}^{(4)} &= \frac{1}{32\pi G} \bar{s}^{\mu\kappa}
    h^{\nu\lambda} {\cal G}_{\mu\nu\kappa\lambda} \,, \label{eq:lagrangian:s}
    \\
    {\cal L}^{(5)}_{\rm SME} &= -\frac{1}{128\pi G} h_{\mu\nu}
    q^{\mu\rho\alpha\nu\beta\sigma\gamma} \partial_\beta
    R_{\rho\alpha\sigma\gamma} \,, \label{eq:lagrangian:q}
\end{align}
where $R_{\rho\alpha\sigma\gamma}$ is the linearized Riemann curvature tensor,
and ${\cal G}_{\mu\nu\kappa\lambda}$ is its double dual; $\bar{s}^{\mu\kappa}$
and $q^{\mu\rho\alpha\nu\beta\sigma\gamma}$ are coefficients for Lorentz/CPT
violation. Components of $\bar{s}^{\mu\kappa}$ are dimensionless, while those
of $q^{\mu\rho\alpha\nu\beta\sigma\gamma}$ have the dimension of the length (or
the inverse mass).  In the operational counting in
SME~\cite{Kostelecky:2003fs}, ${\cal L}_{\rm SME}^{(4)}$ breaks the Lorentz
symmetry, but preserves the CPT symmetry, while ${\cal L}_{\rm SME}^{(5)}$
breaks both Lorentz and CPT symmetries~\cite{Kostelecky:2003fs}.  $\bar
s^{\mu\kappa}$ is a symmetric, traceless tensor, thus it has 9 independent
components.  The first three indices of $q^{\mu\rho\alpha\nu\beta\sigma\gamma}$
are completely antisymmetric, while the last four have the symmetry of the
Riemann tensor.  Thus, there are 60 independent coefficients in
$q^{\mu\rho\alpha\nu\beta\sigma\gamma}$~\cite{Kostelecky:2016kfm,
Bailey:2017lbo}.  Because $\bar{s}^{\mu\kappa}$ has already been discussed in
various literature~\cite{Kostelecky:2003fs, Bailey:2006fd, Kostelecky:2008ts},
we will focus on $q^{\mu\rho\alpha\nu\beta\sigma\gamma}$ in this paper. The
contributions from $\bar{s}^{\mu\kappa}$ are kept in some expressions in the
text, only for interested readers for convenient comparisons; all numerical
calculations in this paper have set $\bar{s}^{\mu\kappa}=0$. As mentioned by
\citet{Bailey:2017lbo}, some specific models have direct or indirect mappings
to the Lagrangian in Eqs.~(\ref{eq:lagrangian:s}) and (\ref{eq:lagrangian:q}),
like the vector field models with a potential term driving spontaneous
Lorentz/diffeomorphism breaking~\cite{Kostelecky:1989jw} and those with
additional beyond-Maxwell kinetic terms~\cite{Jacoby:2006dy}, noncommutative
geometry \cite{Carroll:2001ws}, quantum gravity \cite{Gambini:1998it}, and so
on.

Neglecting higher-order terms, the field equation derived from
Eq.~(\ref{eq:lagrangian}) reads~\cite{Bailey:2017lbo},
\begin{align}
    G^{\mu\nu} = 8\pi G T^{\mu\nu}_{\rm matter} + \bar{s}_{\kappa\lambda}
    {\cal G}^{\mu\kappa\nu\lambda} - \frac{1}{4} q^{\rho\alpha\left( \mu\nu
    \right)\beta\sigma\gamma} \partial_\beta R_{\rho\alpha\sigma\gamma} \,,
\end{align}
where $(\cdot)$ denotes the symmetrization of indices.

With post-Newtonian techniques~\cite{Will:1993ns}, one can derive the
leading-order Lagrangian for two bodies $a$ and $b$~\cite{Bailey:2006fd,
Bailey:2017lbo},
\begin{widetext}
\begin{align}
    L =& \frac{1}{2} \left( m_a v_a^2 + m_b v_b^2 \right) + \frac{G m_a m_b}{r}
    \left( 1 + \frac{3}{2} \bar s_{00} + \frac{1}{2} \bar s_{jk} \hatn^j
    \hatn^k \right) \nonumber \\
    & + \frac{G m_a m_b}{2r} \left[ 3 \bar s_{0j} \left( v_a^j + v_b^j \right)
    + \bar s_{0j} \hatn^j \left( v_a^k + v_b^k \right) \hatn^k \right] -
    \frac{3G m_a m_b}{2r^2} v^j_{ab} \left( K_{jklm} \hatn^k \hatn^l \hatn^m -
    K_{jkkl} \hatn^l \right) \,, \label{eq:2body:lagrangian}
\end{align}
\end{widetext}
where $m_a$ and $m_b$ are masses, $\bm{v}_a$ and $\bm{v}_b$ are velocities (a boldface indicates vectors),
$\bm{r} \equiv \bm{r}_a - \bm{r}_b$ is the relative separation, and
$\hat{\bm{n}} \equiv \bm{r} / r$ with $r \equiv \left| \bm{r} \right|$,
$\bm{v}_{ab} \equiv \bm{v}_a - \bm{v}_b$. 
As can be seen from the second line
of the equation, while the $\bar{s}^{\mu\nu}$ terms depend on the ``absolute''
velocities of bodies, the $K_{jklm}$ terms (to be introduced below) only depend
on the relative velocity of two bodies.  When $\bar{s}^{\mu\nu} = 0$, the
Lagrangian reduces to,
\begin{widetext}
\begin{align}
    L =& \frac{1}{2} \left( m_a v_a^2 + m_b v_b^2 \right) + \frac{G m_a m_b}{r}
    - \frac{3G m_a m_b}{2r^2} v^j_{ab} \left( K_{jklm} \hatn^k \hatn^l \hatn^m
    - K_{jkkl} \hatn^l \right) \,. 
\end{align}
\end{widetext}
In Eq.~(\ref{eq:2body:lagrangian}) we have defined,
\begin{align}\label{eq:K4indices}
    K_{jklm} &\equiv -\frac{1}{6} \left( q_{0jk0l0m} + q_{n0knljm} +
    q_{njknl0m} + \mbox{permutations} \right) \,,
\end{align}
which is the linear combination of $q^{\mu\rho\alpha\nu\beta\sigma\gamma}$ that
enters the post-Newtonian scheme at leading order~\cite{Bailey:2017lbo};
``permutations'' here mean all symmetric permutations in the last three indices
$klm$. While the post-Newtonian limit contains all 9 independent coefficients
in $\bar{s}^{\mu\nu}$, there are only 15  independent combinations of 30
irreducible pieces (out of 60) in $q^{\mu\rho\alpha\nu\beta\sigma\gamma}$
appearing~\cite{Bailey:2017lbo}. This is similar for the Lorentz-violating
effects on the gravitational-wave propagation in SME, where a subset of 16 of
these coefficients appear at leading order~\cite{Kostelecky:2016kfm}.

Using the Euler-Lagrange equation,
\begin{align}
    \frac{d}{dt} \frac{\partial L}{\partial \bm{v}_a} - \frac{\partial
    L}{\partial \bm{r}_a} = 0 \,,
\end{align}
we can obtain from Eq.~(\ref{eq:2body:lagrangian}) the acceleration of
body $a$~\cite{Bailey:2017lbo},
\begin{widetext}
\begin{align}
    \frac{d^2 r_a^j}{dt^2} =& -\frac{Gm_b}{r^2} \left[ \left( 1+\frac{3}{2}
    \bar{s}_{00} \right)  \hatn^j - \bar s_{jk} \hatn^k + \frac{3}{2} \bar
    s_{kl} \hatn^k \hatn^l \hatn^j  \right] + \frac{2Gm_b}{r^2} \left( \bar s_{0j}
    v^k \hatn^k - \bar s_{0k} v^k \hatn^j \right) \nonumber \\
    & + \frac{Gm_b}{r^2} \bar s_{0k} v^l_b \left[ 2 \delta^{j(k} \hatn^{l)} -
    3\delta^{kl}\hatn^j - 3\hatn^j \hatn^k \hatn^l \right] + \frac{G m_b
    v^k}{r^3} \left( 15 \hatn^l \hatn^m \hatn^n \hatn_{[j} K_{k]lmn} + 9
    \hatn^l \hatn^m K_{[jk]lm} - 9 \hatn_{[j} K_{k]llm} \hatn^m - 3K_{[jk]ll}
\right) \,,
\end{align}
\end{widetext}
where $\left[ \cdot \right]$ denotes the anti-symmetrization of indices.  The
acceleration for body $b$ can be obtained by interchanging the body indices $a
\leftrightarrow b$. Again, when $\bar{s}^{\mu\nu} = 0$, the equation reduces
to,
\begin{widetext}
\begin{align}\label{eq:acc:K}
    \frac{d^2 r_a^j}{dt^2} = -\frac{Gm_b}{r^2} \hatn^j + \frac{G m_b v^k}{r^3}
    \left( 15 \hatn^l \hatn^m \hatn^n \hatn_{[j} K_{k]lmn} + 9 \hatn^l \hatn^m
    K_{[jk]lm} - 9 \hatn_{[j} K_{k]llm} \hatn^m - 3K_{[jk]ll} \right) \,.
\end{align}
\end{widetext}
The second term $\propto v/r^3$ of the above equation provides us with a
nonstatic (namely velocity-dependent) inverse cubic force between two masses.
The behaviour of this term is vastly different from what occurs in GR and other
Lorentz-violating terms that preserve the CPT symmetry~\cite{Bailey:2014bta,
Kostelecky:2016uex}.  There is no self-acceleration term in (\ref{eq:acc:K}),
which is consistent with the fact that SME is based on an action principle with
energy and momentum conservation~\cite{Shao:2016ezh}.

\begin{figure}
  \centering
  \includegraphics[width=9cm]{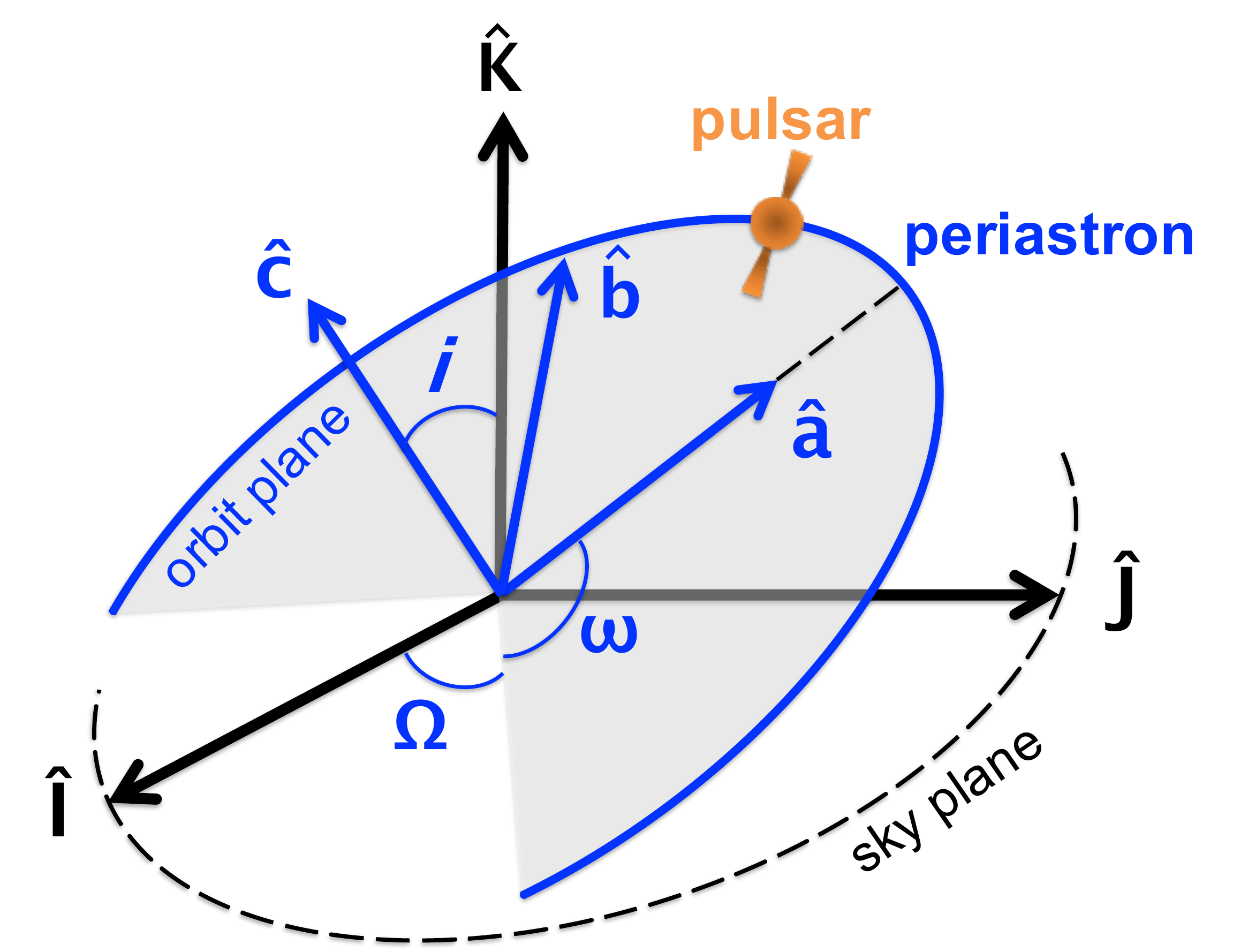}
  \caption{An illustration of coordinate systems~\cite{Shao:2014bfa}. The frame
      $(\hat{\bf I}, \hat{\bf J}, \hat{\bf K})$ is comoving with the pulsar
      system, with $\hat{\bf K}$ pointing along the line of sight to the pulsar
      from the Earth, while $(\hat{\bf I}, \hat{\bf J})$ constitutes the sky
      plane with $\hat{\bf I}$ to east and $\hat{\bf J}$ to north. The spatial
      frame $(\hat{\bf a}, \hat{\bf b}, \hat{\bf c})$ is centered at the pulsar
      system with $\hat{\bf a}$ pointing from the center of mass to the
      periastron, $\hat{\bf c}$ along the orbital angular momentum, and
      $\hat{\bf b} \equiv \hat{\bf c} \times \hat{\bf a}$. The frames,
      $(\hat{\bf I}, \hat{\bf J}, \hat{\bf K})$ and $(\hat{\bf a}, \hat{\bf b},
  \hat{\bf c})$, are related through rotation matrices, ${\cal R}^{(\Omega)}$,
  ${\cal R}^{(i)}$, and ${\cal R}^{(\omega)}$.  \label{fig:orbit}}
\end{figure}

Now we discuss the secular changes for a bound orbit with the accleration
(\ref{eq:acc:K}).  For an elliptical binary orbit, we use the notations in
\citet{Damour:1991rd}. In particular, the coordinate systems $\left( \hat{\bf
I}, \hat{\bf J}, \hat{\bf K} \right)$ and $\left( \hat{\bf a}, \hat{\bf b},
\hat{\bf c} \right)$ are defined in Figure~\ref{fig:orbit}. Notations are the
same as that in Refs.~\cite{Shao:2014bfa, Shao:2014oha}, but differ from
Refs.~\cite{Bailey:2006fd, Bailey:2017lbo} where $\left( \vec{P}, \vec{Q},
\vec{k} \right) \equiv \left( \hat{\bf a}, \hat{\bf b}, \hat{\bf c} \right)$
was used. To connect the spatial frame $\left( \hat{\bf a}, \hat{\bf b},
\hat{\bf c} \right)$ with the cannonical Sun-centered celestial-equatorial
frame, $(\hat{\bf X},\hat{\bf Y},\hat{\bf Z})$, one needs a spatial rotation,
${\cal R}$, to align the axes,\footnote{We neglect the boost between these two
frames, which is small, with $v/c \approx {\cal
O}\left(10^{-3}\right)$, where $v$ is the systematic velocity of the binary
pulsar with respect to the Solar System~\cite{Bailey:2006fd, Shao:2014bfa}.}
\begin{equation}
    \label{eq:rotation:tot}
  \left(
  \begin{array}{c}
    \hat{\bf a} \\
    \hat{\bf b} \\
    \hat{\bf c}
  \end{array}
  \right) =
  {\cal R}
  \left(
  \begin{array}{c}
    \hat{\bf X} \\
    \hat{\bf Y} \\
    \hat{\bf Z}
  \end{array}
  \right) \,.
\end{equation}
With the help of $(\hat{\bf I},\hat{\bf J},\hat{\bf K})$ in
Figure~\ref{fig:orbit}, one can decompose the full rotation into five simple
parts, characterized by parameters in celestial mechanics~\cite{Bailey:2006fd,
Shao:2014bfa, Shao:2014oha},
\begin{equation}
  {\cal R} = {\cal R}^{(\omega)} {\cal R}^{(i)} {\cal R}^{(\Omega)}
  {\cal R}^{(\delta)} {\cal R}^{(\alpha)} \,,
\end{equation}
where
\begin{eqnarray}
  {\cal R}^{(\alpha)} &=&
  \left(
  \begin{array}{ccc}
    -\sin\alpha & \cos\alpha & 0 \\
    -\cos\alpha & -\sin\alpha & 0 \\
    0 & 0 & 1
  \end{array}
  \right) \,,\\
  {\cal R}^{(\delta)} &=&
    \left(
  \begin{array}{ccc}
    1 & 0 & 0 \\
    0 & \sin\delta & \cos\delta \\
    0 & -\cos\delta & \sin\delta
  \end{array}
  \right) \,,\\
  {\cal R}^{(\Omega)} &=&
    \left(
  \begin{array}{ccc}
    \cos\Omega & \sin\Omega & 0 \\
    -\sin\Omega & \cos\Omega & 0 \\
    0 & 0 & 1
  \end{array}
  \right) \,,\\
  {\cal R}^{(i)} &=&
    \left(
  \begin{array}{ccc}
    1 & 0 & 0 \\
    0 & \cos i & \sin i \\
    0 & -\sin i & \cos i
  \end{array}
  \right) \,,\\
  {\cal R}^{(\omega)} &=&
    \left(
  \begin{array}{ccc}
    \cos\omega & \sin\omega & 0 \\
   -\sin\omega & \cos\omega & 0 \\
    0 & 0 & 1
  \end{array}
  \right) \,. \label{eq:rotation:omega}
\end{eqnarray}
In the rotation matrix, $\alpha$ and $\delta$ are the right ascension and
declination of the binary pulsar, $i$ is the orbital inclination, $\omega$ is
the longitude of the periastron, and $\Omega$ is the longitude of the ascending
node (see Figure~\ref{fig:orbit}).

Using the techniques of osculating elements, \citet{Bailey:2017lbo} obtained
the secular changes of orbital elements after averaging over the orbital-period
timescale,
\begin{align}
    \left\langle \frac{da}{dt} \right\rangle &= 0 \,,  \\
    \left\langle \frac{de}{dt} \right\rangle &= 0 \,,  \\
    \label{eq:domdt}
    \left\langle \frac{d\omega}{dt} \right\rangle &= - \frac{n_b^2}{4\left(
    1-e^2 \right)^{3/2}} \left\{ 2 K_1 + \cot i \left[ K_2 \cos\omega + K_3
    \sin \omega \right] \right\} \,, \\
    \label{eq:didt}
    \left\langle \frac{di}{dt} \right\rangle &= \frac{n_b^2}{4\left( 1-e^2
    \right)^{3/2}} \left[ K_3 \cos\omega - K_2 \sin\omega \right] \,, \\
    \left\langle \frac{d\Omega}{dt} \right\rangle &= \frac{n_b^2}{4\left( 1-e^2
    \right)^{3/2}} \csc i \left[ K_2 \cos\omega + K_3 \sin\omega \right] \,,
\end{align}
where $a$ is the semimajor axis, $e$ is the orbital eccentricity, and $n_b
\equiv 2\pi/P_b$ with $P_b$ the orbital period. In above equations, $K_1$,
$K_2$, $K_3$ are defined by~\cite{Bailey:2017lbo},
\begin{align}
    \label{eq:K1}
  K_1 &\equiv 3 K_{\hata\hata\hata\hatb} + K_{\hata\hatb\hatb\hatb} + 6
  K_{[\hata\hatb]\hatc\hatc} \,, \\
    \label{eq:K2}
  K_2 &\equiv 3 K_{\hata\hatb\hatb\hatc} - 3 K_{\hata\hata\hata\hatc} - 4
  K_{\hata\hatc\hatc\hatc} - 6 K_{\hatb\hata\hatb\hatc} \,, \\
    \label{eq:K3}
  K_3 &\equiv 6 K_{\hata\hata\hatb\hatc} + 4 K_{\hatb\hatc\hatc\hatc} -
  3K_{\hatb\hata\hata\hatc} + 3 K_{\hatb\hatb\hatb\hatc} \,,
\end{align}
where the indices on the right hand sides denote the projection of $K_{jklm}$
in Eq.~(\ref{eq:K4indices}) onto the $\left( \hat{\bf a}, \hat{\bf b}, \hat{\bf
c} \right)$ directions. More details can be found in
Ref.~\cite{Bailey:2017lbo}.

\bgroup
\def\arraystretch{1.25}
\begin{table*}
    \caption{\label{tab:dns}
	Relevant timing parameters for
	PSRs~B1913+16~\cite{Weisberg:2016jye}, B1534+12~\cite{Fonseca:2014qla},
	B2127+11C~\cite{Jacoby:2006dy}, and J0737$-$3039A~\cite{Kramer:2006nb}.
	Parenthesized numbers represent the 1-$\sigma$ uncertainty in the last
	digits quoted. Estimated parameters are marked with ``$\spadesuit$''.}
	\begin{tabular}{p{5.5cm}p{2.85cm}p{2.85cm}p{2.85cm}p{2.85cm}}
	\hline\hline
	& PSR~B1913+16 & PSR~B1534+12 & PSR~B2127+11C & PSR~J0737$-$3039A \\
	\hline
	Observational span, $T_{\rm obs}$ (year) & $\sim31$ & $\sim22$ &
	$\sim12$ & $\sim2.7$ \\
	Right ascension, $\alpha$ (J2000) &  \JRA{19}{15}{27}{99942(3)} &
	\JRA{15}{37}{09}{961730(3)} & \JRA{21}{30}{01}{2042(1)} &
	\JRA{07}{37}{51}{24927(3)} \\
	Declination, $\delta$ (J2000) &  \Jdec{16}{06}{27}{3868(5)} &
	\Jdec{11}{55}{55}{43387(6)} & \Jdec{12}{10}{38}{209(4)} &
	\Jdec{-30}{39}{40}{7195(5)} \\
	Orbital period, $P_b$ (day) & $0.322997448918(3)$ & $0.420737298879(2)$
	& $0.33528204828(5)$ & $0.10225156248(5)$ \\
	Eccentricity, $e$ & $0.6171340(4)$ & $0.27367752(7)$ & $0.681395(2)$ &
	$0.0877775(9)$ \\
	Pulsar's projected semimajor axis, $x_p$ (lt-s) & $2.341776(2)$ &
	$3.7294636(6)$ & $2.51845(6)$ & $1.415032(1)$ \\
	Longitude of periastron, $\omega$ (deg) & $292.54450(8)$ &
	$283.306012(12)$ & $345.3069(5)$ & $87.0331(8)$ \\
	Epoch of periastron, $T_0$ (MJD) & $52144.90097849(3)$ &
	$52076.827113263(11)$ & $50000.0643452(3)$ & $53155.9074280(2)$ \\
	Advance of periastron, $\dot\omega$ (deg\,yr$^{-1}$) & $4.226585(4)$ &
	 $1.7557950(19)$ & $4.4644(1)$ & $16.89947(68)$ \\
	 Time derivative of $x_p$, $\dot x_p$ & $-1.4(9)\times10^{-14}$ &
	 $\left|\dot x_p\right| < 3.0 \times 10^{-15} \, \spadesuit$ & $\left|
	 \dot x_p \right| < 5.5 \times 10^{-13} \, \spadesuit$ & $\left|\dot
	 x_p\right| < 4.1 \times 10^{-14} \, \spadesuit$ \\
	Parameters used to derive masses & $\gamma$ \& $\dot P_b$ & $\gamma$ \&
	$s$ & $\gamma$ \& $\dot P_b$ & $R$ \& $s$ \\
	Pulsar mass, $m_1$ ($M_\odot$) & $1.435(2)$ & $1.364(20)$ & $1.36(4)$ &
	$1.339(3)$ \\
	Companion mass, $m_2$ ($M_\odot$) & $1.390(1)$ & $1.356(7)$ & $1.36(2)$
	& $1.250(2)$ \\
	Excess of $\dot\omega$, $\dot\omega - \dot\omega^{\rm GR}$
	(deg\,yr$^{-1}$) & $0.003(3)$ & $-0.018(12)$ & $0.00(6)$ & $-0.01(2)$
	\\
	\hline
    \end{tabular}
\end{table*}
\egroup

\bgroup
\def\arraystretch{1.25}
\begin{table*}
    \caption{\label{tab:nswd:optical}
	Relevant timing parameters for
	PSRs~J0348+0432~\cite{Antoniadis:2013pzd},
	J1738+0333~\cite{Freire:2012mg}, and
	J1012+5307~\cite{Lazaridis:2009kq}.  Parenthesized numbers represent
	the 1-$\sigma$ uncertainty in the last digits quoted. The listed
	Laplace-Lagrange parameter, $\eta$, is the intrinsic value, after
	subtraction of the contribution from the Shapiro
	delay~\cite{Lange:2001rn}.  Masses are derived from the combination of
	optical and radio observations, and they are independent of the
        underlying gravity theory~\cite{Wex:2014nva, Shao:2016ezh}. Estimated
        parameters are marked with ``$\spadesuit$''.}
    \begin{tabular}{p{6.5cm}p{3.5cm}p{3.5cm}p{3.5cm}}
	\hline\hline
	& PSR J0348+0432 & PSR J1738+0333 & PSR J1012+5307 \\
	\hline
	Observational span, $T_{\rm obs}$ (year) & $\sim3.7$ & $\sim10.0$ &
	$\sim15.0$ \\
	Right ascension, $\alpha$ (J2000) & \JRA{03}{48}{43}{639000(4)} &
	\JRA{17}{38}{53}{9658386(7)} & \JRA{10}{12}{33}{4341010(99)} \\
	Declination, $\delta$ (J2000) & \Jdec{04}{32}{11}{4580(2)} &
	\Jdec{03}{33}{10}{86667(3)} & \Jdec{53}{07}{02}{60070(13)} \\
	Orbital period, $P_b$ (day) & $0.102424062722(7)$ &
	$0.3547907398724(13)$ & $0.60467271355(3)$ \\
	Pulsar's projected semimajor axis, $x_p$ (lt-s) & $0.14097938(7)$ &
	$0.343429130(17)$ & $0.5818172(2)$ \\
	$\eta \equiv e\sin\omega$ & $\left(1.9 \pm 1.0\right) \times 10^{-6}$ &
	$\left( -1.4\pm1.1 \right)\times10^{-7}$ & $\left( -1.4\pm3.4
	\right)\times10^{-7}$ \\
	$\kappa \equiv e\cos\omega$ & $\left( 1.4\pm1.0 \right) \times 10^{-6}$
	& $\left( 3.1\pm1.1 \right) \times 10^{-7}$ & $\left( 0.6\pm3.1
	\right)\times10^{-7}$ \\
	Time derivative of $x_p$, $\dot x_p$ &  $\left| \dot x_p\right| <
	2.1\times10^{-15} \, \spadesuit$ & $\left( 0.7\pm0.5
	\right)\times10^{-15}$ & $\left( 2.3\pm0.8 \right) \times10^{-15}$ \\
	Pulsar mass, $m_1$ ($M_\odot$) & $2.01(4)$ & $1.46^{+0.06}_{-0.05}$ &
	$1.64(22)$ \\
	Companion mass, $m_2$ ($M_\odot$) & $0.172(3)$ &
	$0.181^{+0.008}_{-0.007}$ & $0.16(2)$ \\
	\hline
    \end{tabular}
\end{table*}
\egroup

\bgroup
\def\arraystretch{1.25}
\begin{table*}
    \caption{\label{tab:nswd:nooptical}
	Relevant timing parameters for
	PSRs~J0751$+$1807~\cite{Desvignes:2016yex},
	J1802$-$2124~\cite{Ferdman:2010rk},
	J1909$-$3744~\cite{Desvignes:2016yex}, and
	J2043+1711~\cite{Arzoumanian:2017puf}.  Parenthesized numbers represent
    the 1-$\sigma$ uncertainty in the last digits quoted.  Estimated parameters
    are marked with ``$\spadesuit$''.}
	\begin{tabular}{p{5.5cm}p{2.85cm}p{2.85cm}p{2.85cm}p{2.85cm}}
	\hline\hline
	& PSR~J0751+1807 & PSR~J1802$-$2124 & PSR~J1909$-$3744 & PSR~J2043+1711
	\\
	\hline
	Observational span, $T_{\rm obs}$ (year) & $\sim17.6$ & $\sim6.4$ &
	$\sim9.4$ & $\sim4.5$ \\
	Right ascension, $\alpha$ (J2000) & \JRA{07}{51}{09}{155331(13)} &
	\JRA{18}{02}{05}{335576(5)} & \JRA{19}{09}{47}{4335737(7)} &
	\JRA{20}{43}{20}{881730(1)}  \\
	Declination, $\delta$ (J2000) & \Jdec{18}{07}{38}{4864(10)} &
	\Jdec{-21}{24}{03}{649(2)} & \Jdec{-37}{44}{14}{51561(3)} &
	\Jdec{17}{11}{28}{91265(3)} \\
	Orbital period, $P_b$ (day) & $0.263144270792(7)$ & $0.698889243381(5)$
	& $1.533449474329(13)$ & $1.482290786394(15)$ \\
	Pulsar's projected semimajor axis, $x_p$ (lt-s) & $0.3966158(3)$ &
	$3.7188533(5)$ & $1.89799099(6)$ & $1.62395834(15)$ \\
	$\eta \equiv e\sin\omega$ & $\left( 3.3\pm0.5 \right)\times10^{-6}$ &
	$\left( 8.6\pm0.9 \right)\times10^{-7}$ & $\left( 0\pm1.9
	\right)\times10^{-8}$ & $\left( -4.07\pm0.07 \right)\times10^{-6}$ \\
	$\kappa \equiv e\cos\omega$ & $\left( 3.8\pm5.0 \right)\times10^{-7}$ &
	$\left( 2.32 \pm 0.04 \right)\times10^{-6}$ & $\left(- 1.22\pm0.11
	\right)\times10^{-7}$ & $\left( -2.67\pm0.05 \right)\times10^{-6}$ \\
	Time derivative of $x_p$, $\dot x_p$ & $\left( -4.9\pm0.9
	\right)\times10^{-15}$ & $\left|\dot x_p \right| < 8.5 \times
	10^{-15}\,\spadesuit$ & $\left( 0.6\pm1.7 \right)\times10^{-16}$ &
	$\left|\dot x_p \right| < 3.7 \times 10^{-15}\,\spadesuit$\\
	Parameters used to derive masses & $\dot P_b$ \& $\zeta$ & $r$ \& $s$ &
	$r$ \& $s$ & $h_3$ \& $\zeta$ \\
	Pulsar mass, $m_1$ ($M_\odot$) & $1.64(15)$ & $1.24(11)$ & $1.540(27)$
	& $1.38^{+0.12}_{-0.13}$ \\
	Companion mass, $m_2$ ($M_\odot$) & $0.16(1)$ & $0.78(4)$ &
	$0.2130(24)$ & $0.173(10)$ \\
	\hline
    \end{tabular}
\end{table*}
\egroup

\section{Binary pulsars}
\label{sec:psr}

Our starting point to put constraints on the SME coefficients with binary
pulsars will be using the secular changes in orbital elements.  In general,
pulsar timing is insensitive to the longitude of the ascending node $\Omega$,
unless the binary is very nearby~\cite{1996ApJ...467L..93K, Lorimer:2005}.
Thus, the secular changes in the orbital inclination and the longitude of the
periastron are the most relevant to our tests. A nonzero $\left\langle di/dt
\right\rangle$ will be reflected in the accurately measured, projected
semimajor axis of the pulsar orbit, $x_p \equiv a_p \sin i/c$, where $a_p
\simeq m_2 a / \left( m_1+m_2 \right)$ is the semimajor axis of the pulsar
orbit.\footnote{We hereafter use $m_1$ and $m_2$ to denote the masses of the
pulsar and its companion, respectively.}  From Eq.~(\ref{eq:didt}), one has,
\begin{align}\label{eq:dxpdt}
    \left\langle \frac{\dot x_p}{x_p} \right\rangle = \frac{ n_b^2
    \cot i}{4\left( 1-e^2 \right)^{3/2}} \left[ K_3 \cos\omega - K_2 \sin\omega
    \right] \,.
\end{align}

In the following, we will make use of Eqs.~(\ref{eq:domdt}) and
(\ref{eq:dxpdt}), naming them as the $\dot\omega$-test and the $\dot x_p$-test
respectively, to put bounds on the coefficients for Lorentz/CPT violation.  It
is apparent from Eqs.~(\ref{eq:domdt}) and (\ref{eq:dxpdt}) that binary pulsars
with small orbits will provide tight constraints.  Besides the smallness of the
orbit, there are other criteria to meet for binary pulsars, that will become
clear later.  According to the needs for the $\dot\omega$-test and/or the $\dot
x_p$-test, we carefully pick 11 well-timed binary pulsars with relativistic
orbits. We categorize them into three groups:
\begin{enumerate}
    \item Group I: relativistic double NS binaries with orbital
	period smaller than 1 day. We pick 4 binary pulsars:
	PSRs~B1913+16~\cite{Weisberg:2016jye}, B1534+12~\cite{Fonseca:2014qla},
	B2127+11C~\cite{Jacoby:2006dy}, and J0737$-$3039A~\cite{Kramer:2006nb}.
	Relevant timing parameters for our tests are listed in
	Table~\ref{tab:dns}.
      \item Group II: relativistic neutron-star--white-dwarf (NS-WD) binaries with
	orbital period smaller than 1 day, and whose WD companions
	were well studied with optical observations.  We pick 3 binary
	pulsars: PSRs~J0348+0432~\cite{Antoniadis:2013pzd},
	J1738+0333~\cite{Freire:2012mg}, and
	J1012+5307~\cite{Lazaridis:2009kq}.  Relevant timing parameters for
	our tests are listed in Table~\ref{tab:nswd:optical}.
    \item Group III: relativistic NS-WD binaries with
	orbital period smaller than 2 days, and whose Shapiro delays were
	also identified in the timing observations. We pick 4 binary
	pulsars: PSRs~J0751$+$1807~\cite{Desvignes:2016yex},
	J1802$-$2124~\cite{Ferdman:2010rk},
	J1909$-$3744~\cite{Desvignes:2016yex}, and
	J2043+1711~\cite{Arzoumanian:2017puf}.  Relevant timing parameters for
	our tests are listed in Table~\ref{tab:nswd:nooptical}.
\end{enumerate}

\bgroup
\def\arraystretch{1.25}
\begin{table}
    \caption{Constraints on $K_i ~ (i=1,2,3)$ from binary pulsars. Notice that
    the definition of $K_i$ depends on the geometry of the binary through
    projections in Eqs.~(\ref{eq:K1}--\ref{eq:K3}). 
 \label{tab:Ki}}
    \begin{tabular}{p{2.6cm}p{0.9cm}p{4.8cm}}
	\hline\hline
	Pulsar & Test & 1-$\sigma$ constraint \\
	\hline
	PSR~J0348+0432 & $\dot x_p$ & $\left| 0.81 K_2 - 0.59 K_3  \right| <
	30\,{\rm m}$ \\
	PSR~J0737$-$3039A & $\dot x_p$ & $\left| 0.99K_2 - 0.13 K_3 \right| <
	2.0\,{\rm km}$ \\
	& $\dot \omega$ & $\left|2K_1 + 0.03 K_3 \right| < 26 \,{\rm km}$ \\
	PSR J0751+1807 & $\dot x_p$ & $\left| 0.99K_2 - 0.11K_3 \right| <
	81\,{\rm m}$ \\
	PSR J1012+5307 & $\dot x_p$ & $\left|0.92 K_2 +  0.39 K_3 \right| <
	140\,{\rm m}$ \\
	PSR~B1534+12 & $\dot x_p$ & $\left| 0.97 K_2 + 0.24 K_3 \right| <
	132\,{\rm m}$ \\
	& $\dot\omega$ & $\left| 2K_1 + 0.05 K_2 - 0.21 K_3 \right| < 240 \,
	{\rm km}$ \\
	PSR J1738+0333 & $\dot x_p$ & $\left|0.41 K_2  + 0.91 K_3 \right| <
	27\,{\rm m}$\\
	PSR J1802$-$2124 & $\dot x_p$ & $\left| 0.35K_2 - 0.94K_3\right| <
	1.8\,{\rm km}$ \\
	PSR J1909$-$3744 & $\dot x_p$ & $\left| K_3 \right| < 670 \,{\rm m} $\\
	PSR~B1913+16 & $\dot x_p$ & $\left|0.99 K_2 - 0.16 K_3  \right| <
	48\,{\rm m}$ \\
	 & $\dot\omega$ & $\left| 2K_1 - 0.15 K_2 - 0.92 K_3 \right| < 19 \,
	{\rm km}$ \\
	PSR J2043+1711 & $\dot x_p$ & $\left| 0.84K_2 - 0.55K_3 \right| < 8.6
	\,{\rm km}$ \\
	PSR B2127+11C & $\dot x_p$ & $\left| 0.29 K_2 + 0.96 K_3 \right| <
	2.6\,{\rm km}$ \\
	& $\dot\omega$ & $\left| 2K_1 + 0.80 K_2 - 0.25 K_3 \right| < 330
	\,{\rm km}$ \\
	\hline
    \end{tabular}
\end{table}
\egroup

\bgroup
\def\arraystretch{1.25}
\begin{table*}
    \caption{\label{tab:single}
    Limits on different components of $q^{\mu\rho\alpha\nu\beta\sigma\gamma}$,
assuming only one of them is nonzero. Components $q^{\rm XYZXYZT}$ and $q^{\rm
XYZXZYT}$ do not enter the tests from binary pulsars, thus they remain
unconstrained.}
\centering
\begin{tabular}{p{2.2cm}p{2.4cm}p{2.2cm}p{2.4cm}p{2.2cm}p{2.4cm}}
	\hline\hline
	Coefficient & 1-$\sigma$ limit [m] &
	Coefficient & 1-$\sigma$ limit [m] &
	Coefficient & 1-$\sigma$ limit [m]
	\\
	\hline
	$q^{\rm TXYTXTX}$ & 22 &      
	$q^{\rm TXYTXTY}$ & 11 &      
	$q^{\rm TXYTXTZ}$ & 12 \\     
	$q^{\rm TXYTYTY}$ & 10 &      
	$q^{\rm TXYTYTZ}$ & 5.7 &      
	$q^{\rm TXYTZTZ}$ & 9.7 \\      
	$q^{\rm TXYXYXY}$ & 8.0 &      
	$q^{\rm TXYXYXZ}$ & 8.3 &      
	$q^{\rm TXYXYYZ}$ & 6.2 \\      
	$q^{\rm TXYXZXZ}$ & 8.3 &      
	$q^{\rm TXYXZYZ}$ & 3.7 &      
	$q^{\rm TXYYZYZ}$ & 5.3 \\      
	$q^{\rm TXZTXTX}$ & 24 &      
	$q^{\rm TXZTXTY}$ & 10 &      
	$q^{\rm TXZTXTZ}$ & 11 \\      
	$q^{\rm TXZTYTY}$ & 6.2 &      
	$q^{\rm TXZTYTZ}$ & 4.8 &      
	$q^{\rm TXZTZTZ}$ & 18 \\      
	$q^{\rm TXZXZXZ}$ & 27 &      
	$q^{\rm TXZXZYZ}$ & 11 &      
	$q^{\rm TXZYZYZ}$ & 6.5 \\      
	$q^{\rm TYZYZYZ}$ & 8.8 &      
	$q^{\rm XYZXYXT}$ & 29 &      
	$q^{\rm XYZXYYT}$ & 14 \\      
	$q^{\rm XYZXYZT}$ & --- &      
	$q^{\rm XYZXZXT}$ & 13 &      
	$q^{\rm XYZXZYT}$ & --- \\      
	$q^{\rm XYZXZZT}$ & 14 &      
	$q^{\rm XYZYZYT}$ & 13 &      
	$q^{\rm XYZYZZT}$ & 29 \\
	\hline
    \end{tabular}
\end{table*}
\egroup

These 11 binary pulsars all have been monitored for years, most of which were
regularly observed within the pulsar-timing-array projects, including the Parks
Pulsar Timing Array (PPTA) \cite{Hobbs:2013aka}, the European Pulsar Timing
Array (EPTA) \cite{Kramer:2013kea}, and the North American Nanohertz
Observatory for Gravitational Waves (NANOGrav) \cite{McLaughlin:2013ira}.  To
successfully achieve the proposed $\dot\omega$-test and/or $\dot x_p$-test, we
address the following concerns: 
\begin{itemize}
    \item Because $\dot x_p$ was not always fitted for in deriving the timing
	solution of binary pulsars, wherever it is inaccessible, we
	conservatively estimate a 1-$\sigma$ upper limit from the uncertainty
	of $x_p$, as $\left| \dot x_p \right|^{\rm upper} = \sqrt{12}
	\sigma_{x_p} / T_{\rm obs}$~\cite{Shao:2014oha}, where $T_{\rm obs}$ is
	the time span used in deriving the timing solution.  The prefactor
	``$\sqrt{12}$'' was inspired by a linear-in-time evolution.  Actually
	as was already noticed for PSR~B1534+12, this is a quite good
	estimation \cite{Shao:2014oha}. In addition, PSR~B1913+16 was estimated
	by \citet{Shao:2014oha} to have $\left| \dot x_p \right|^{\rm upper} =
	1.3 \times 10^{-14}$ using the results of \citet{Weisberg:2010zz} where
	$\dot x_p$ was not reported. Recently, \citet{Weisberg:2016jye} fitted
	for $\dot x_p$, and obtained $\dot x_p = \left( -1.4 \pm 0.9 \right)
	\times 10^{-14}$ in excellent agreement with the estimation. This
	further gives us confidence in using the estimation formula. Estimated
	$\dot x_p$'s are decorated with ``$\spadesuit$'' in
	Tables~\ref{tab:dns}, \ref{tab:nswd:optical}, and
	\ref{tab:nswd:nooptical}.
    \item Sometimes for nearby binary pulsars, there is a contribution to $\dot
	x_p$ from the proper motion of the binary~\cite{1996ApJ...467L..93K},
	\begin{align}\label{eq:PM} \left(\frac{\dot x_p}{x_p}\right)^{\rm PM} =
	    \left( -\mu_\alpha \sin\Omega + \mu_\delta \cos\Omega \right) \cot
	i \,, \end{align} where $\mu_\alpha$ and $\mu_\delta$ are proper
	motions in $\alpha$ and $\delta$ directions
	respectively~\cite{Lorimer:2005}.  It could produce a nonzero $\dot
	x_p$, as was measured for several binary pulsars.  Assuming GR as the
	theory of gravity, this piece of information can be used to constrain
	$\Omega$.  Here we do not assume GR and stay agnostic about the
	longitude of ascending node. We randomly distribute it
	uniformly in the range $\Omega \in [0, 360^\circ)$; thus the net effect
	from Eq.~(\ref{eq:PM}) after averaging over $\Omega$ vanishes. For
	these pulsars with reported $\dot x_p$'s, we take the uncertainty of
	the observed $\dot x_p$ as an estimate for its upper limit.
    \item Usually, for double NS binaries in Group I, the total mass
	of the binary is calculated from the very well measured
	$\dot\omega$~\cite{Lorimer:2005}.  For consistency, the
	$\dot\omega$-test is invalid if masses were derived from the observed
	$\dot\omega$ by assuming GR.  Therefore, we need to re-calculate masses
	without using the measured $\dot\omega$. We performed such calculations
	for PSRs~B1913+16, B1534+12, B2127+11C, and J0737$-$3039A.  Results are
	listed in Table~\ref{tab:dns}.  By using these $\dot\omega$-independent
	masses, we recalculate the periastron advance rate with GR, and obtain
	the excess of $\dot\omega$ by substracting it from the observed value.
	By doing so, we obtain a ``clean'' $\dot\omega$-test.  The
	uncertainties in the excess of $\dot\omega$ are dominated by the
	uncertainties of the masses, and as a cost the clean $\dot\omega$-test
	usually gives much worse limits than those from $\dot x_p$ (see
	Table~\ref{tab:Ki}). This will be the bottleneck for our global
	analysis (see below).
    \item One caution in directly using the secular change of $\omega$ in
	Lorentz-violating theories was pointed out by \citet{Wex:2007ct}, that
	a large $\dot\omega$ can render the secular changes nonconstant. These
	effects cannot be too large based on the fact that all binaries were
	well fitted with simple timing models. In our samples, the biggest
	change in $\omega$ is $\sim100^\circ$ for PSR~B1913+16
	\cite{Weisberg:2016jye}. Therefore, we consider it safe to use
	time-averaged values for $\omega$-related quantities as a rough
	approximation at current stage.\footnote{This will not be valid for
	    the (unpublished) new timing solution of the double
	pulsar~\cite{Kramer:2006nb, Kramer:2016kwa} where, assuming GR, up to
    now a change in $\omega$ is $> 250^\circ$ already.} For example, in
    Eqs.~(\ref{eq:domdt}) and (\ref{eq:dxpdt}), we use the $\omega$ value in
    the middle of the observational span. In principle, a timing model with
    nonlinear-in-time evolution of $\omega$ would be perfect in addressing this
    issue~\cite{Wex:2007ct}, which is rather complicated and it is beyond the
    scope of this work (see Ref.~\cite{Wex:2007ct} for a simplified version
    when assuming an edge-on orbit, approximating the double pulsar).
    \item As was pointed out several
        times, $\Omega$ is in general not determined in pulsar timing. We will
        treat it a random variable uniformly distributed in $\Omega \in [0,
    360^\circ)$. This choice makes our tests ``probabilistic tests''.
    \item To perform the $\dot\omega$-test and the $\dot x_p$-test, component
	masses of the binary are needed sometimes. We have discussed the
	situation for double NS binaries in Group I. For NS-WD 
	binaries in Group II, we use the masses derived from the
	optical observation of the WD. These masses are independent of
	the gravity theories~\cite{Wex:2014nva, Shao:2016ezh} (see
	Table~\ref{tab:nswd:optical}). For NS-WD binaries
	in Group III, we derive masses from the measurement of the Shapiro
	delay for PSRs~J1802$-$2124, J1909$-$3744, and J2043+1711, while for
	PSR~J0751+1807, we also used the orbital decay measurement for
	assistance (see Table~\ref{tab:nswd:nooptical}). These calculation
	assumes that the deviations from GR are small, in consistent with the
	observational results, as well as the effective-field-theory
	framework. Nevertheless, we might overlook strong-field effects that
	arise in some specific theories~\cite{Damour:1993hw, Sennett:2017lcx,
	Shao:2017gwu} (see section~\ref{sec:diss}).
\end{itemize}

\bgroup
\def\arraystretch{2}
\begin{table*}
    \caption{\label{tab:global}
    Global constraints on the canonical set of 15 $K_{jklm}$.}
\centering
\begin{tabular}{p{2.4cm}p{11cm}p{2.4cm}}
	\hline\hline
	Symbol & Definition & 1-$\sigma$ limit [$10^6$\,m]
	\\
	\hline
	$K_{\rm XXXY}$ & $\frac{1}{3} \left( -q^{\rm TXYTXTX} + q^{\rm TXYXYXY}
	+ q^{\rm TXYXZXZ} - q^{\rm XYZXZXT} \right)$ & 6.6 \\
	$K_{\rm XXXZ}$ & $\frac{1}{3} \left( q^{\rm TXYXYXZ} - q^{\rm TXZTXTX}
	+ q^{\rm TXZXZXZ} + q^{\rm XYZXYXT} \right)$ & 3.1 \\
	$K_{\rm XXYY}$ & $\frac{1}{3} \left( -2 q^{\rm TXYTXTY} +2 q^{\rm
	TXYXZYZ} + q^{\rm XYZXYZT} -2 q^{\rm XYZXZYT} \right)$ & 7.1 \\ 
	$K_{\rm XXYZ}$ & $\frac{1}{6} \left( -2 q^{\rm TXYTXTZ} -2 q^{\rm
	TXYXYYZ} - 2 q^{\rm TXZTXTY} + 2 q^{\rm TXZXZYZ} + q^{\rm XYZXYYT}
        -q^{\rm XYZXZZT} \right)$ & 2.7 \\
	$K_{\rm XXZZ}$ & $\frac{1}{3} \left( -2 q^{\rm TXYXZYZ} -2 q^{\rm
	TXZTXTZ} + 2 q^{\rm XYZXYZT} -q^{\rm XYZXZYT} \right)$ & 8.1 \\
	$K_{\rm XYYY}$ & $-q^{\rm TXYTYTY} +q^{\rm TXYXYXY} + q^{\rm TXYYZYZ} -
	q^{\rm XYZYZYT}$ & 20 \\
	$K_{\rm XYYZ}$ & $\frac{1}{3} \left( -2 q^{\rm TXYTYTZ} +3 q^{\rm
	TXYXYXZ} -q^{\rm TXZTYTY} + q^{\rm TXZYZYZ} -q^{\rm XYZYZZT} \right)$
	& 3.1 \\
	$K_{\rm XYZZ}$ & $\frac{1}{3} \left( -q^{\rm TXYTZTZ} + 3 q^{\rm
	TXYXZXZ} + q^{\rm TXYYZYZ} - 2 q^{\rm TXZTYTZ} - q^{\rm XYZYZYT}
        \right)$ & 6.6 \\
	$K_{\rm XZZZ}$ & $-q^{\rm TXZTZTZ} + q^{\rm TXZXZXZ} + q^{\rm TXZYZYZ}
	- q^{\rm XYZYZZT}$ & 9.3 \\
	$K_{\rm YXXZ}$ & $\frac{1}{3} \left( 3 q^{\rm TXYTXTZ} + 3 q^{\rm
	TXYXYYZ} - q^{\rm TXZTXTY} + q^{\rm TXZXZYZ} + q^{\rm XYZXZZT} \right)$
	& 2.7 \\
	$K_{\rm YXYZ}$ & $\frac{1}{6} \left( 4 q^{\rm TXYTYTZ} -2 q^{\rm
	TXYXYXZ} - 2 q^{\rm TXZTYTY} + 2 q^{\rm TXZYZYZ} + q^{\rm XYZXYXT} +
        q^{\rm XYZYZZT} \right)$ & 3.1 \\
	$K_{\rm YXZZ}$ & $\frac{1}{3} \left( 3 q^{\rm TXYTZTZ} - q^{\rm
	TXYXZXZ} - 3 q^{\rm TXYYZYZ} - 2 q^{\rm TXZTYTZ} + q^{\rm XYZXZXT}
        \right)$ & 6.6 \\
	$K_{\rm YYYZ}$ & $\frac{1}{3} \left( q^{\rm TXYXYYZ} - q^{\rm TXZTYTY}
	+ q^{\rm TYZYZYZ} + q^{\rm XYZXYYT} \right)$ & 2.7 \\
	$K_{\rm YYZZ}$ & $\frac{1}{3}\left( 2 q^{\rm TXYXZYZ} - 2 q^{\rm
	TXZTYTZ} + q^{\rm XYZXYZT} +q^{\rm XYZXZYT} \right)$ & 4.0 \\
	$K_{\rm YZZZ}$ & $-q^{\rm TXZTZTZ} + q^{\rm TXZXZYZ} + q^{\rm TYZYZYZ}
	+ q^{\rm XYZXZZT}$ & 8.0 \\
	\hline
    \end{tabular}
\end{table*}
\egroup

\begin{figure*}
    \includegraphics[width=18cm]{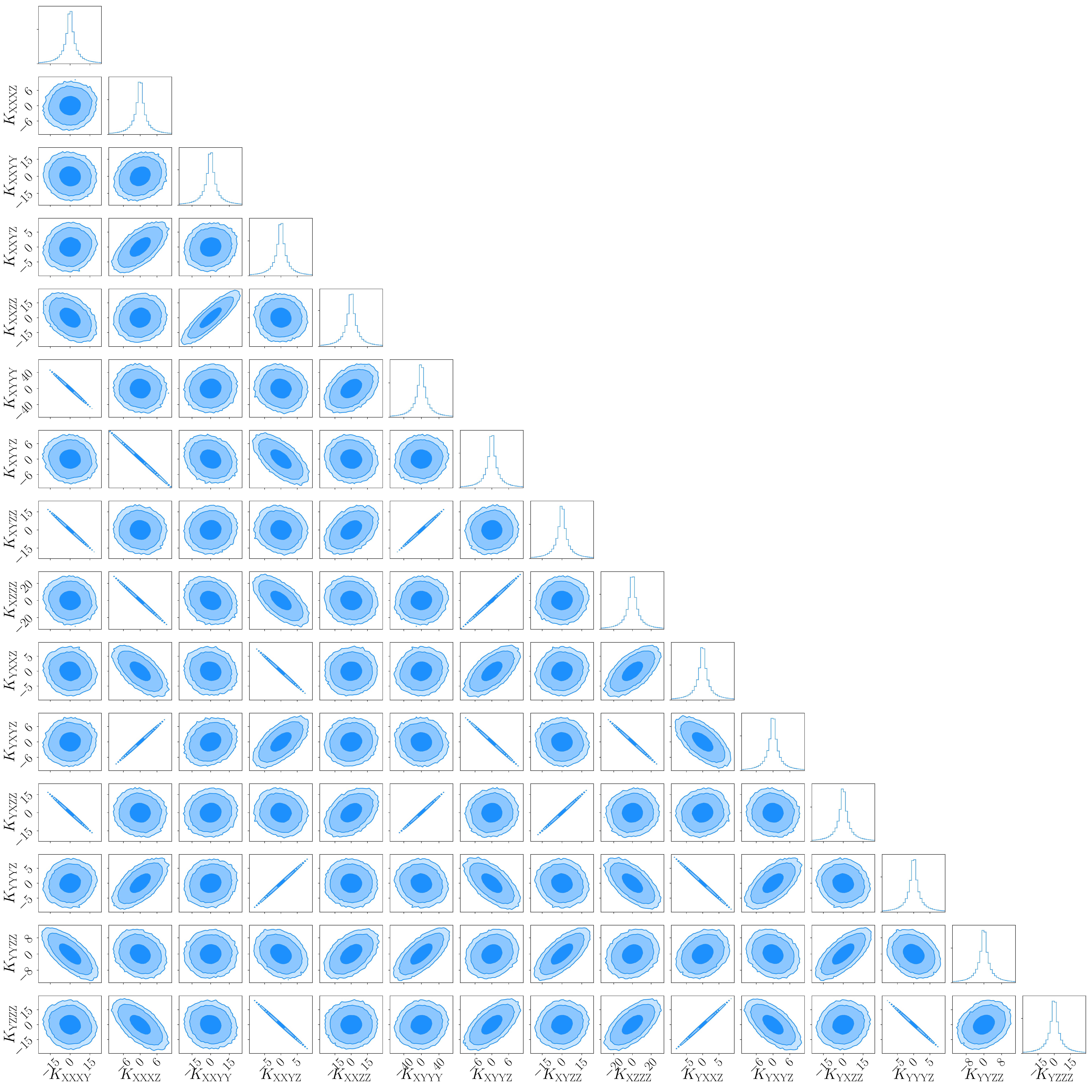}
    \caption{\label{fig:Ks} Contours and histograms of the set of 15
    independent $K_{jklm}$'s in our simulation. Contours show the 68\%,  90\%,
and 95\% confidence levels.  The unit for $K_{jklm}$ is $10^6\,$m in this
figure.}
\end{figure*}

Taking the above considerations into account, we have derived a set of
independent limits on various linear combinations of coefficients for
Lorentz/CPT violation, making use of 4 $\dot\omega$-tests and 11 $\dot
x_p$-tests from the pulsars in Tables~\ref{tab:dns}, \ref{tab:nswd:optical},
and \ref{tab:nswd:nooptical}.  These results are tabulated in
Table~\ref{tab:Ki}, and the best ones are in agreement with the estimation by
\citet{Bailey:2017lbo}.  Notice that, the results in Table~\ref{tab:Ki} should
be directly compared with the {\it estimated sensitivity} in the Table~1 of
Ref.~\cite{Bailey:2017lbo}. The estimated sensitivities for other experiments,
namely the Solar system ephemeris, laser ranging, gravimeter, short-range
gravity, and time delay, are expected to be orders of magnitude weaker.
Nevertheless, it would still be valuable to work out the actual limits that
these experiments would cast; they might probe some components of
$q^{\mu\rho\alpha\nu\beta\sigma\gamma}$ which binary pulsars are insensitive to
study (see below).

The limits in Table~\ref{tab:Ki} are not of fundamental value. $K_i$'s
$(i=1,2,3)$ are system dependent through the projections defined in
Eqs.~(\ref{eq:K1}--\ref{eq:K3}), where projections are given explicitly in
Eqs.~(\ref{eq:rotation:tot}--\ref{eq:rotation:omega}) with various angles
different for individual pulsars. This is the power of many pulsar systems that
are in principle able to break any parameter degeneracy~\cite{Shao:2014oha,
Shao:2014bfa}.  In order to convert the limits in Table~\ref{tab:Ki} into
limits on the underlying Lorentz-violating coefficients
$q^{\mu\rho\alpha\nu\beta\sigma\gamma}$ in the
Lagrangian~(\ref{eq:lagrangian:q}), we use Eq.~(\ref{eq:K4indices}) to relate
$K_{ijlm}$ with $q^{\mu\rho\alpha\nu\beta\sigma\gamma}$. 

The limits in Table~\ref{tab:Ki} are limits on different linear combinations of
$q^{\mu\rho\alpha\nu\beta\sigma\gamma}$. For a 1-$\sigma$ limit ``$a$'', we
denote it as $\left| {\cal X}_a \left(  q^{\mu\rho\alpha\nu\beta\sigma\gamma},
\Omega_a \right)\right| < {\cal C}_a$ where the longitude of the ascending node
$\Omega_a$ is unknown in general. To proceed practically, we adopt the
probabilistic density function,
\begin{widetext}
\begin{align} \label{eq:prob}
    P\left( q^{\mu\rho\alpha\nu\beta\sigma\gamma}\right) \propto\prod_a
    \int_0^{2\pi} \frac{1}{\sqrt{2\pi}}\exp \left\{- \frac{1}{2} \left|\frac{
    {\cal X}_a \left( q^{\mu\rho\alpha\nu\beta\sigma\gamma}, \Omega_a
    \right)}{{\cal C}_a} \right|^2 \right\} \frac{d\Omega_a}{2\pi} \,,
\end{align}
\end{widetext}
where we have made assumptions on the Gaussianity of measurements and that the
limits on $K_i$'s in Table~\ref{tab:Ki} are mutually independent. In
Eq.~(\ref{eq:prob}) we have also marginalized over the unknwon angles
$\Omega_a$, as a {\it nuisance} parameter in the language of Bayesian
statistics~\cite{Patrignani:2016xqp}.

As mentioned in section~\ref{sec:theory}, from Young tableaux it was
established that there are 60 independent coefficients for
$q^{\mu\rho\alpha\nu\beta\sigma\gamma}$, while only a subset of 30 (in the form
of 15 independent linear combinations) could appear in our pulsar
tests~\cite{Bailey:2017lbo}. We identify them explicitly. We find that,
actually 2 of these 30 coefficients, $q^{\rm XYZXYZT}$ and $q^{\rm XYZXZYT}$,
do not show up. This phenomenon was already met in other contexts of SME
\cite{Kostelecky:2008in}. It tells us that binary-pulsar tests will not be able
to constrain these 2 components, and even if they are large, they can escape
from our tests. They need to be constrained with other experiments.  The
conclusion is worked out through an explicit calculation, but we do not have a
clear physical understanding why this particular set of coefficients are
relevant to binary pulsars.  However, relaxing the assumptions  (i.e.,
post-Newtonian order ${\cal O}(v/c)$ beyond the Newtonian limit) and using
spin-weighted spherical harmonics could reveal more precisely the underlying
reasons for the combinations of coefficients appearing in this analysis
\cite{Kostelecky:2007fx, Kostelecky:2016uex, Kostelecky:2009zp}. We hope it
stimulates other groups to analyze their experiments, and obtain a better
understanding.

As a first attempt to constrain $q^{\mu\rho\alpha\nu\beta\sigma\gamma}$, we
treat only one of them as nonzero. The final limit comes from a properly
weighted combination of the 15 tests in Table~\ref{tab:Ki}. The results are
listed in Table~\ref{tab:single}.  In the scenario where only one of
$q^{\mu\rho\alpha\nu\beta\sigma\gamma}$ is nonzero, the constraint is derived
predominantly from the tightest ones in Table~\ref{tab:Ki}. The coefficients
for Lorentz/CPT violation $q^{\mu\rho\alpha\nu\beta\sigma\gamma}$ are limited
to ${\cal O}\left( 1\mbox{--}10\,{\rm m} \right)$, as predicted by
\citet{Bailey:2017lbo}.

In addition, we perform a global test where all 15 independent combinations of
$q^{\mu\rho\alpha\nu\beta\sigma\gamma}$ could be nonzero. In this case, we use
a set of 15 canonical $K_{jklm}$ to represent these linear combinations. They
are identified explicitly and are given in terms of
$q^{\mu\rho\alpha\nu\beta\sigma\gamma}$ in the second column of
Table~\ref{tab:global}. Since we have 15 independent terms, we have to use all
15 tests given in Table~\ref{tab:Ki}. As was done for $\bar s^{\mu\nu}$ in
Ref.~\cite{Shao:2014oha}, Monte Carlo simulations are set up to properly
account for the measurements and the unknown $\Omega$'s. Our results are given
in Figure~\ref{fig:Ks}, and the marginalized distributions are utilized to
derive the 1-$\sigma$ constraints on the set of 15 canonical $K_{jklm}$, and
they are given in the last column of Table~\ref{tab:global}. In this scenario
we are only able to constrain $K_{jklm}$ to the level ${\cal O}\left( 10^6
\,{\rm m} \right)$.  The direct limits in Table~\ref{tab:Ki} are quite
heteroscedastic, spanning from ${\cal O}\left( 10\,{\rm m} \right)$ to ${\cal
O}\left( 10^5\,{\rm m} \right)$. Because of this, the global analysis gives
limits corresponding {\it more or less} to the worst limits in
Table~\ref{tab:Ki} with strong correlations between some coefficients (see
Figure~\ref{fig:Ks}). In future, more tests will tighten these limits.

Our results in Tables~\ref{tab:single} and \ref{tab:global} constitute the
first set of systematic limits from pulsar timing experiments on
$q^{\mu\rho\alpha\nu\beta\sigma\gamma}$. They are also the first set of
constraints from the post-Newtonian dynamics of binaries with CPT-violating
operators in SME for the gravity sector, complementary to the unique limit
obtained from the kinematics in the propagation of gravitational
waves~\cite{Kostelecky:2016kfm}.  Because the SME is viewed as an {\it
effective field theory}, the coefficients for Lorentz/CPT violation are not
fixed {\it a priori} \cite{Kostelecky:2003fs, Bailey:2006fd}.  In general,
specific theories are needed to cast predictions for their values. We here
undertake an agnostic way, and let data decide the values they can have and the
constraints they should satisfy. Our results can be mapped to theory parameters
if a theory is specified.
 
\section{Discussions} \label{sec:diss}

Searching for new physics beyond the current paradigm is a rewarding task.  Up
to now, no violation in Lorentz and CPT symmetries has been convincingly found
\cite{Kostelecky:2008ts, Will:2014kxa, Shao:2016ezh}.  When the deviation is
perturbatively small, the effective-field-theory framework of SME provides a
practically useful platform to systematically study these tiny deviations.
Many new phenomena were discovered in SME for the past decades. Here we
specifically study the pure gravity sector of SME in the presence of
CPT-violating leading-order operators with mass dimension 5. These operators
are interesting in the following manner. While being of higher mass dimension
than those of GR and  the leading-order Lorentz-violating operators which are
of mass dimension 4, they can be better probed with astronomical observations
instead of short-range experiments~\cite{Bailey:2017lbo}. The insensitivity of
short-range laboratory experiments is due to the nature of CPT violation where,
an additional suppression factor, proportional to $\left( \bm{v}_a - \bm{v}_b
\right) / c$, is present. In order to confine short-range experiments within
laboratories for a long duration for precision measurement, this factor appears
enormously small.  In contrast, for relativistic binary pulsars with $P_b
\lesssim 1$\,day, this factor can be as large as $10^{-3}$. Therefore, binary
pulsars become even more powerful than short-range experiments to constrain
these operators.

Motivated by this observation, in this paper we have utilized binary pulsars to
constrain these operators, using the analytical post-Newtonian results for a
binary orbit from \citet{Bailey:2017lbo}. By taking care of all caveats from
observational facts, we tailored the results into a form that can immediately
be used in analysing binary pulsars. Well-timed relativistic binary pulsars
turn out to be suitable for the tests, and we put constraints on the
coefficients for Lorentz/CPT violation to ${\cal O}\left( 10 \,{\rm m}\right)$
when only one coefficient is allowed to be nonzero (see
Table~\ref{tab:single}), and to ${\cal O}\left( 10^6\,{\rm m} \right)$ when all
coefficients can be nonzero at the same time (see Table~\ref{tab:global}). They
represent the first set of observational constraints for CPT-violating
gravity in SME from the post-Newtonian dynamics, complementing the kinematic
constraints from gravitational waves~\cite{Kostelecky:2016kfm}. 

Since the SME is based on the perturbative nature of effective field
theories~\cite{Kostelecky:2003fs} and in particular here we have used the
linearized gravity~\cite{Bailey:2017lbo, Kostelecky:2017zob}, our limits on
$q^{\mu\rho\alpha\nu\beta\sigma\gamma}$ cannot probe nonperturbative effects
that {\it might} arise with the strong gravitational fields of NSs, like
the ``scalarization'' phenomenon in scalar-tensor theories~\cite{Damour:1993hw,
Freire:2012mg, Shao:2017gwu, Sennett:2017lcx}. Strictly speaking, our limits
are {\it effective} limits for the strong-field counterparts of
$q^{\mu\rho\alpha\nu\beta\sigma\gamma}$. Nevertheless, usually the strong-field
limits are more restricting than their weak-field counterparts. Thus, our
results are actually conservative in this respect. 
The constancy of $q^{\mu\rho\alpha\nu\beta\sigma\gamma}$ in our
  work is an assumption that is required for the energy-momentum
  conservation of the Lagrangian \cite{Kostelecky:2003fs}.  It does
  not leave out the possibility of variations in these coefficients on
  timescales longer than those in which the Sun-centered frame is
  approximately inertial, i.e., a few hundreds of years. In more
  general cases, for example, when considering the strong-field
  effects from NSs, one {\it might} get body-dependent, or in some
  cases even position-dependent coefficients for Lorentz/CPT violation
  (e.g., a term similar to the Whitehead's term in the parameterized
post-Newtonian framework \cite{Will:1973zz, Gibbons:2006jy,
Shao:2013eka, Shao:2017lmp}).  But this will need some specific
theoretic inputs and is beyond the scope of this work.

Pulsar timing in the future will further improve the measurements of binary
orbits, and provide better limits on possible new physics. In our case, the
measurement precisions for $\dot\omega$ and $\dot x_p$ both improve as
$T^{-3/2}$ \cite{Damour:1991rd} where $T$ is the observational time span, even
without improvements in the telescopes.  Nevertheless, we in addition have new
telescopes and technologies coming online. The upcoming observations at the
Five-hundred-meter Aperture Spherical Telescope (FAST)~\cite{Nan:2011um} and
the Square Kilometre Array (SKA)~\cite{Shao:2014wja, Kramer:2004hd} are guaranteed
to boost the timing precision. Also, they will discover more binary pulsars to
perform the tests. Therefore, the actual improvement in constraining the
coefficients for Lorentz/CPT violation will be significantly faster than $T^{-3/2}$.

\acknowledgments
We thank Alan Kosteleck\'y and Norbert Wex for helpful discussions, and Paulo
Freire for carefully reading the manuscript. We also thank the anonymous
referee for helpful comments.  This work was supported by the National Science
Foundation of China (11721303), and XDB23010200. LS acknowledges financial
support by the European Research Council (ERC) for the ERC Synergy Grant
BlackHoleCam under Contract No. 610058, and is grateful to the Mainz Institute
for Theoretical Physics (MITP) for its hospitality and its partial support
during the completion of this work.

\end{document}